\documentstyle[11pt,psfig,twoside]{article}

\begin{document}

\newcommand{\dd}{\,{\rm d}}
\newcommand{\ie}{{\it i.e.},\,}
\newcommand{\etal}{{\it et al.\ }}
\newcommand{\eg}{{\it e.g.},\,}
\newcommand{\cf}{{\it cf.\ }}
\newcommand{\vs}{{\it vs.\ }}
\newcommand{\zdot}{\makebox[0pt][l]{.}}
\newcommand{\up}[1]{\ifmmode^{\rm #1}\else$^{\rm #1}$\fi}
\newcommand{\dn}[1]{\ifmmode_{\rm #1}\else$_{\rm #1}$\fi}
\newcommand{\upd}{\up{d}}
\newcommand{\uph}{\up{h}}
\newcommand{\upm}{\up{m}}
\newcommand{\ups}{\up{s}}
\newcommand{\arcd}{\ifmmode^{\circ}\else$^{\circ}$\fi}
\newcommand{\arcm}{\ifmmode{'}\else$'$\fi}
\newcommand{\arcs}{\ifmmode{''}\else$''$\fi}
\newcommand{\MS}{{\rm M}\ifmmode_{\odot}\else$_{\odot}$\fi}
\newcommand{\RS}{{\rm R}\ifmmode_{\odot}\else$_{\odot}$\fi}
\newcommand{\LS}{{\rm L}\ifmmode_{\odot}\else$_{\odot}$\fi}

\newcommand{\Abstract}[2]{{\footnotesize\begin{center}ABSTRACT\end{center}
\vspace{1mm}\par#1\par
\noindent
{~}{\it #2}}}

\newcommand{\TabCap}[2]{\begin{center}\parbox[t]{#1}{\begin{center}
  \small {\spaceskip 2pt plus 1pt minus 1pt T a b l e}
  \refstepcounter{table}\thetable \\[2mm]
  \footnotesize #2 \end{center}}\end{center}}

\newcommand{\TableSep}[2]{\begin{table}[p]\vspace{#1}
\TabCap{#2}\end{table}}

\newcommand{\FigCap}[1]{\footnotesize\par\noindent Fig.\  %
  \refstepcounter{figure}\thefigure. #1\par}

\newcommand{\TableFont}{\footnotesize}
\newcommand{\TableFontIt}{\ttit}
\newcommand{\SetTableFont}[1]{\renewcommand{\TableFont}{#1}}

\newcommand{\MakeTable}[4]{\begin{table}[htb]\TabCap{#2}{#3}
  \begin{center} \TableFont \begin{tabular}{#1} #4 
  \end{tabular}\end{center}\end{table}}

\newcommand{\MakeTableSep}[4]{\begin{table}[p]\TabCap{#2}{#3}
  \begin{center} \TableFont \begin{tabular}{#1} #4 
  \end{tabular}\end{center}\end{table}}

\newenvironment{references}%
{
\footnotesize \frenchspacing
\renewcommand{\thesection}{}
\renewcommand{\in}{{\rm in }}
\renewcommand{\AA}{Astron.\ Astrophys.}
\newcommand{\AAS}{Astron.~Astrophys.~Suppl.~Ser.}
\newcommand{\ApJ}{Astrophys.\ J.}
\newcommand{\ApJS}{Astrophys.\ J.~Suppl.~Ser.}
\newcommand{\ApJL}{Astrophys.\ J.~Letters}
\newcommand{\AJ}{Astron.\ J.}
\newcommand{\IBVS}{IBVS}
\newcommand{\PASP}{P.A.S.P.}
\newcommand{\Acta}{Acta Astron.}
\newcommand{\MNRAS}{MNRAS}
\renewcommand{\and}{{\rm and }}
\section{{\rm REFERENCES}}
\sloppy \hyphenpenalty10000
\begin{list}{}{\leftmargin1cm\listparindent-1cm
\itemindent\listparindent\parsep0pt\itemsep0pt}}%
{\end{list}\vspace{2mm}}

\def\TYLDA{~}
\newlength{\DW}
\settowidth{\DW}{0}
\newcommand{\dw}{\hspace{\DW}}

\newcommand{\refitem}[5]{\item[]{#1} #2%
\def\REFARG{#3}\ifx\REFARG\TYLDA\else, {\it#3}\fi
\def\REFARG{#4}\ifx\REFARG\TYLDA\else, {\bf#4}\fi
\def\REFARG{#5}\ifx\REFARG\TYLDA\else, {#5}\fi.}

\newcommand{\Section}[1]{\section{#1}}
\newcommand{\Subsection}[1]{\subsection{#1}}
\newcommand{\Acknow}[1]{\par\vspace{5mm}{\bf Acknowledgements.} #1}
\pagestyle{myheadings}

\def\thefootnote{\fnsymbol{footnote}}
\begin{center}
{\large\bf The Optical Gravitational Lensing Experiment.\\
%\vskip3pt
Cepheids in the Magellanic Clouds.\\
%\vskip3pt
IV.  Catalog of Cepheids from the Large Magellanic Cloud\footnote
{Based on  observations obtained with the 1.3~m Warsaw telescope at the
Las Campanas  Observatory of the Carnegie Institution of Washington.}}
\vskip0.8cm
{\bf
A.~~U~d~a~l~s~k~i$^1$,~~I.~~S~o~s~z~y~{\'n}~s~k~i$^1$,
~~M.~~S~z~y~m~a~{\'n}~s~k~i$^1$,
~~M.~~K~u~b~i~a~k$^1$,~~G.~~P~i~e~t~r~z~y~\'n~s~k~i$^1$,
~~P.~~W~o~\'z~n~i~a~k$^2$,~~ and~~K.~~\.Z~e~b~r~u~\'n$^1$}
\vskip3mm
{$^1$Warsaw University Observatory, Al.~Ujazdowskie~4, 00-478~Warszawa, Poland\\
e-mail: (udalski,soszynsk,msz,mk,pietrzyn,zebrun)@sirius.astrouw.edu.pl\\
$^2$ Princeton University Observatory, Princeton, NJ 08544-1001, USA\\
e-mail: wozniak@astro.princeton.edu}
\end{center}

\Abstract{
We present the Catalog of Cepheids from the LMC. The Catalog contains
1333 objects detected in the 4.5 square degree area of central parts of
the LMC. About $3.4\cdot10^5$ {\it BVI} measurements of these stars were
collected during the OGLE-II microlensing survey. The Catalog data
include period, {\it BVI} photometry, astrometry, and $R_{21},
\phi_{21}$ parameters of the Fourier decomposition of {\it I}-band light
curve.

The vast majority of objects from the Catalog are the classical Cepheids
pulsating in the fundamental or first overtone mode. The remaining
objects include Population II Cepheids and red giants with pulsation-like
light curves.

Tests of completeness performed in overlapping parts of adjacent fields
indicate that completeness of the Catalog is very high: $>96$\%.
Statistics and distributions of basic parameters of Cepheids are also
presented. 

Finally, we show the light curves of three eclipsing systems containing
Cepheid detected among objects of the Catalog.

All presented data, including individual {\it BVI} observations are
available from the OGLE Internet archive.
}{~}

\Section{Introduction}

Cepheids were among the first variable stars discovered by astronomers
-- the prototype of the class, $\delta$ Cep, and $\eta$ Aql were found
to be varying in brightness by J.\ Goodricke and E.\ Pigott,
respectively,  in 1784. The great career of these objects begun at the
beginning of 20th century when their famous Period--Luminosity relation
was discovered in the Small Magellanic Cloud by Leavitt (1912). Cepheids
became one of the most important standard candles used for distance
determination in the Universe, although the calibration of the
Period--Luminosity relation is still a topic of lively debate. Proper
calibration is of great importance because Cepheids are now routinely
discovered in galaxies to about 25 Mpc with the HST instruments.
Thus, the Cepheid based distance scale is one of the most important
steps in the distance scale ladder.

Cepheids are relatively well understood pulsating stars. Their role in
the modern astrophysics is hard to be overestimated. Beside the
Period--Luminosity relation these objects are the ideal laboratory for
testing the stellar structure, theory of stellar evolution etc.
Therefore it is crucial to have at hand a large sample of these stars
with good quality observational data so the theoretical work could be
verified.

Although many Cepheids were discovered in the Galaxy their observational
data are very inhomogeneous, taken by different observers with different
instruments. Two nearby galaxies -- the Magellanic Clouds -- are
potentially much better hosts of these objects. Both Large and Small
Magellanic Clouds are known to contain many Cepheids. Additional
advantage of Cepheids from these galaxies is that they are located at
approximately the same distance what makes analyses of their properties
much simpler.

Unfortunately both Magellanic Clouds have been neglected photometrically
for years -- the vast majority of known Cepheids in the Magellanic
Clouds were observed with old photographic or photoelectric techniques
giving an order of magnitude worse quality than the modern CCD-based
techniques. Situation has  significantly changed in 1990s when large
microlensing surveys begun regular monitoring of the Magellanic Clouds
for microlensing events. Photometry of millions stars in both Magellanic
Clouds is a natural by-product of these surveys and for the first time
good quality light curves of the Magellanic Cloud Cepheids could be
obtained. Both MACHO and EROS microlensing surveys reported discovery of
many Cepheids and presented observations of these stars in the LMC
and/or SMC (Alcock \etal 1995, Alcock \etal 1999, Sasselov \etal 1997,
Bauer \etal 1999). Unfortunately, all these data were taken with
non-standard photometric bands.

The Magellanic Clouds were added to the list of objects observed by the
Optical Gravitational Lensing Experiment (OGLE) at the beginning of the
second phase of the survey in January 1997. Since then both Magellanic
Clouds are monitored regularly, practically on every clear night.
Observations are made through the {\it BVI} filters very closely
reproducing the standard {\it BVI} system. After more than two years of
observations the photometric databases are complete enough so the search
for variable stars could be performed. Large samples of Cepheids were
extracted from databases of both Magellanic Cloud fields.

In the previous papers of this series we presented analysis of
double-mode Cepheids in the SMC (Udalski \etal 1999a), discovery of 13
Cepheids in the SMC -- candidates for objects pulsating in the second
overtone mode (Udalski \etal 1999b) and analysis of the
Period--Luminosity and Period--Luminosity--Color relations of huge
samples of Cepheids from the LMC and SMC (Udalski \etal 1999c). In this
paper we present first of two  Catalogs of Cepheids from the Magellanic
Clouds -- the Catalog of Cepheids from the LMC. Similar Catalog of
about 2300 Cepheids from the SMC will follow.

The Catalog of Cepheids from the LMC contains 1333 objects. They come
from the 4.5 square degree area of central parts of the LMC. The vast
majority of them are the classical Cepheids. Other stars include a
sample of Population II Cepheids and a sample of red giant objects which
variability resembles pulsation-like light curves. We do not include
additional sample of about 70 double-mode Cepheids detected in the LMC
-- these objects will be described in a separate paper similar to 
double-mode Cepheids from the SMC (Udalski \etal 1999a).

We also present statistics and distributions of basic parameters of the
LMC Cepheids like location in the LMC, periods, colors and parameters of
the Fourier decomposition of light curve. Finally, we point attention to
three Cepheids in the eclipsing systems which could potentially provide
precise data on sizes and masses of their components.

The large and homogeneous sample of Cepheids presented in this paper
with high quality photometry and high completeness can be used for many
projects concerning these stars. Therefore we decided to make the data
public -- all data presented in this paper, including individual {\it
BVI}-band observations and finding charts are available from the OGLE
Internet archive.

\Section{Observations}

All observations presented in this paper were carried out during the
second  phase of the OGLE experiment with the 1.3-m Warsaw telescope at
the Las  Campanas Observatory, Chile, which is operated by the Carnegie
Institution of  Washington. The telescope was equipped with the "first
generation" camera with  a SITe ${2048\times2048}$ CCD detector working
in drift-scan mode. The  pixel size was 24~$\mu$m giving the 0.417
arcsec/pixel scale. Observations of  the LMC were performed in the
"slow" reading mode of CCD detector with the  gain 3.8~e$^-$/ADU and
readout noise of about 5.4~e$^-$. Details of the  instrumentation setup
can be found in Udalski, Kubiak and Szyma{\'n}ski  (1997). 

\MakeTable{lcc}{12.5cm}{Equatorial coordinates of the OGLE-II LMC fields}
{
\hline
\noalign{\vskip3pt}
\multicolumn{1}{c}{Field} & RA (J2000)  & DEC (J2000)\\
\hline
\noalign{\vskip3pt}
LMC$\_$SC1  &  5\uph33\upm49\ups & $-70\arcd06\arcm10\arcs$ \\
LMC$\_$SC2  &  5\uph31\upm17\ups & $-69\arcd51\arcm55\arcs$ \\
LMC$\_$SC3  &  5\uph28\upm48\ups & $-69\arcd48\arcm05\arcs$ \\
LMC$\_$SC4  &  5\uph26\upm18\ups & $-69\arcd48\arcm05\arcs$ \\
LMC$\_$SC5  &  5\uph23\upm48\ups & $-69\arcd41\arcm05\arcs$ \\
LMC$\_$SC6  &  5\uph21\upm18\ups & $-69\arcd37\arcm10\arcs$ \\
LMC$\_$SC7  &  5\uph18\upm48\ups & $-69\arcd24\arcm10\arcs$ \\
LMC$\_$SC8  &  5\uph16\upm18\ups & $-69\arcd19\arcm15\arcs$ \\
LMC$\_$SC9  &  5\uph13\upm48\ups & $-69\arcd14\arcm05\arcs$ \\
LMC$\_$SC10 &  5\uph11\upm16\ups & $-69\arcd09\arcm15\arcs$ \\
LMC$\_$SC11 &  5\uph08\upm41\ups & $-69\arcd10\arcm05\arcs$ \\
LMC$\_$SC12 &  5\uph06\upm16\ups & $-69\arcd38\arcm20\arcs$ \\
LMC$\_$SC13 &  5\uph06\upm14\ups & $-68\arcd43\arcm30\arcs$ \\
LMC$\_$SC14 &  5\uph03\upm49\ups & $-69\arcd04\arcm45\arcs$ \\
LMC$\_$SC15 &  5\uph01\upm17\ups & $-69\arcd04\arcm45\arcs$ \\
LMC$\_$SC16 &  5\uph36\upm18\ups & $-70\arcd09\arcm40\arcs$ \\
LMC$\_$SC17 &  5\uph38\upm48\ups & $-70\arcd16\arcm45\arcs$ \\
LMC$\_$SC18 &  5\uph41\upm18\ups & $-70\arcd24\arcm50\arcs$ \\
LMC$\_$SC19 &  5\uph43\upm48\ups & $-70\arcd34\arcm45\arcs$ \\
LMC$\_$SC20 &  5\uph46\upm18\ups & $-70\arcd44\arcm50\arcs$ \\
LMC$\_$SC21 &  5\uph21\upm14\ups & $-70\arcd33\arcm20\arcs$ \\
\hline}

Observations of the LMC started on January~6, 1997. 11 driftscan fields 
covering $14.2\times 57$~arcmins in the sky were observed during the
first months of 1997. Additional 10 fields were added in October 1997
increasing the total observed area of the LMC to about 4.5 square degree.
Because the microlensing search  is planned to last for several years,
observations of selected fields will be  continued during the following
seasons.  In this paper we present data  collected up to June 1999.

Observations were obtained in the standard {\it  BVI}-bands with
majority of measurements made in the {\it I}-band. The effective
exposure time was 125, 174 and 237 seconds for the {\it I, V} and {\it
B}-band, respectively. The instrumental system closely resembles the
standard {\it BVI} one -- the color coefficients of transformation
($a\cdot CI$; $a$ -- color coefficient, $CI$ -- color index: $B-V$ for
{\it B} and $V-I$ for {\it VI} filters) are equal to $-0.041$, $+0.004$
and $+0.032$ for the {\it B, V} and {\it I}-band, respectively. 

Due to microlensing search observing strategy the vast majority of
observations were done through the {\it I}-band filter (about $120-360$
epochs depending on the field) while images on about $15-40$ epochs were
collected in the {\it BV}-bands. The {\it B}-band photometry is at the
writing of this paper less complete than {\it VI} photometry --
reductions of only 40\% of fields were finished. For the remaining
fields only {\it VI} photometry was available. {\it B}-band photometry
of these fields will be completed after the next observing season.

Collected images were reduced with the standard OGLE data pipeline.
Quality of data is similar to the photometric data of the SMC described
in Udalski \etal (1998b). In particular, accuracy of absolute photometry
zero points is about $0.01-0.02$ in all {\it BVI}-bands. More details on
the LMC photometric data will be presented with release of the
photometric maps of the LMC in the near future.

\begin{figure}[htb]
\vspace*{10.5cm}
\FigCap{OGLE-II fields in the LMC. Dots indicate positions of Cepheids
from the Catalog. North is up and East to the left in the Digitized Sky
Survey image of the LMC.}
\end{figure}

Table~1 lists equatorial coordinates of center of each field and its
acronym. Fig.~1 shows the Digitized Sky Survey image of the LMC with
contours of the observed fields.

\Section{Selection of Cepheids}

The search for variable objects in the LMC fields was performed using 
observations in the {\it I}-band in which most observations were 
obtained. Typically about 120--360 epochs were available for each
analyzed  object with the lower limit set to~50. The mean {\it I}-band
magnitude of analyzed objects was limited to ${I<19.5}$~mag. Candidates
for variable stars were  selected based on comparison of the standard
deviation of all individual  measurements of a star with typical
standard deviation for stars of similar  brightness. Light curves of
selected candidates were then searched for periodicity using the AoV
algorithm (Schwarzenberg-Czerny 1989). The period search was limited to
the range of 0.1--100~days. Accuracy of periods is about
${7\cdot10^{-5}\cdot P}$.

Candidates for Cepheids were selected from the entire sample of variable
stars  based on visual inspection of the light curves and location in
the color-magnitude diagram (CMD) within the area limited by
${I{<}18.5}$~mag and  ${0.25{<}(V{-}I){<}1.3}$~mag. Several objects
located outside this region (\eg highly  reddened Cepheids) and objects
with no color information but with evident Cepheid-type light curves
were also included to this sample. In total more than 1500 Cepheid
candidates were found in the 4.5 square degree area of the LMC center. 

Each of the analyzed LMC fields overlaps with neighboring fields for 
calibration purposes. Therefore several dozen Cepheids located in the 
overlapping regions were detected twice. We decided not to remove them
from the final list of objects because their measurements are
independent in both  fields and can be used for testing quality of data,
completeness of the sample etc. 105 such objects were detected and we
provide cross-reference list to identify them.

\Section{Basic Parameters of Candidates}

\Subsection{Intensity Mean Photometry}

For each object which passed our selection criteria we derived the {\it
BVI} intensity mean photometry by integrating the light curve converted
to intensity units. It was  approximated by the Fourier series of fifth
order. Result was converted back to the magnitude scale. Accuracy of the
mean {\it I}-band photometry is about $0.001-0.005$~mag and somewhat
worse (about 0.01~mag) for poorer sampled {\it BV}-bands. 

Full {\it BVI} photometry is available only for eight fields:
LMC$\_$SC1--LMC$\_$SC8. For the remaining fields the {\it B}-band
databases are not complete enough for precise determination of the mean
brightness. Photometry of these fields will be completed after the next
observing season of the LMC.

For each object we also determined the extinction insensitive index
$W_I$ (called also Wesenheit index, Madore and Freedman 1991):

$$ W_I=I-1.55*(V-I) \eqno{(1)}  $$

The coefficient 1.55 in Eq.~1 corresponds to the coefficient resulting
from standard interstellar extinction curve dependence of the {\it
I}-band extinction on $E(V-I)$ reddening (\eg Schlegel, Finkbeiner and
Davis 1998). It is easy to show that the values of $W_I$  are the same
when derived from observed or extinction free magnitudes, provided that
extinction to the object is not too high so it can be approximated with
a linear function of color.

\Subsection{Interstellar Reddening}

Determination of the interstellar reddening  to the LMC Cepheids has an
important role in analyses of these objects, distance determination etc.
It is well known that the reddening in the LMC is clumpy and variable
(Harris, Zaritsky and Thompson 1998), therefore applying the mean
reddening value to all objects is generally not justified.

With large photometric databases of millions stars we are in position to
determine the average reddening in many lines-of-sight within the LMC.
Unfortunately, we do not have {\it U}-band photometry which would allow
to derive the reddening from young, hot OB stars. Therefore we used for
this purpose much older but much more numerous red clump stars. It
should be noted, however, that Cepheid population can be distributed in
the LMC somewhat differently than red clump stars and OB-stars
determination could be more appropriate for Cepheids. On the other hand
the differences should not be large for the LMC seen almost face-on.

We used red clump stars for mapping the fluctuations of mean reddening
in our observed fields treating their mean {\it I}-band magnitude as the
reference brightness. It was shown to be independent on age of these
stars in the wide range of $2-10$~Gyr, and it is only slightly dependent
on metallicity (Udalski 1998a,b). The latter correction is not important
in this case because of practically homogeneous environment of field
stars in the LMC (Bica \etal 1998). Thus, the mean brightness of red
clump stars can be a very good reference of brightness for monitoring
extinction. Similar method was used by Stanek (1996) for determination
of extinction map of Baade's Window in the Galactic bulge.

The reddening in the LMC was determined in 84 lines-of-sight.  We
divided each of our 21 $2048\times8192$ pixel fields to four
$2048\times2048$ pixel subfields (subfield 1: $0<y<2048$, etc.). In each
of the subfields we determined the mean observed {\it I}-band magnitude
of red clump stars with technique identical to that described in Udalski
\etal (1998a). Differences of the observed {\it I}-band magnitudes were
assumed as differences of the mean $A_I$ extinction. We converted
differences of $A_I$ extinction to differences of $E(B-V)$ reddening
assuming the standard extinction curve: $E(B-V)=A_I/1.96$ (Schlegel
\etal 1998).

\MakeTable{lcccc}{12.5cm}{$E(B-V)$ reddening in the LMC fields.}
{\hline
\multicolumn{1}{c}{Field}& Subfield~1   & Subfield~2  & Subfield~3  & Subfield~4\\
            & $E(B-V)$     & $E(B-V)$    & $E(B-V)$    & $E(B-V)$\\
\hline
LMC$\_$SC1 & 0.117 & 0.152 & 0.147 & 0.163 \\
LMC$\_$SC2 & 0.121 & 0.121 & 0.150 & 0.131 \\
LMC$\_$SC3 & 0.134 & 0.120 & 0.123 & 0.117 \\
LMC$\_$SC4 & 0.130 & 0.120 & 0.105 & 0.118 \\
LMC$\_$SC5 & 0.130 & 0.115 & 0.108 & 0.133 \\
LMC$\_$SC6 & 0.138 & 0.125 & 0.107 & 0.123 \\
LMC$\_$SC7 & 0.143 & 0.138 & 0.142 & 0.146 \\
LMC$\_$SC8 & 0.131 & 0.133 & 0.136 & 0.142 \\
LMC$\_$SC9 & 0.143 & 0.165 & 0.156 & 0.149 \\
LMC$\_$SC10& 0.156 & 0.147 & 0.146 & 0.132 \\
LMC$\_$SC11& 0.147 & 0.154 & 0.150 & 0.152 \\
LMC$\_$SC12& 0.152 & 0.146 & 0.127 & 0.139 \\
LMC$\_$SC13& 0.154 & 0.129 & 0.135 & 0.130 \\
LMC$\_$SC14& 0.124 & 0.142 & 0.138 & 0.127 \\
LMC$\_$SC15& 0.145 & 0.125 & 0.147 & 0.126 \\
LMC$\_$SC16& 0.135 & 0.148 & 0.185 & 0.181 \\
LMC$\_$SC17& 0.171 & 0.193 & 0.175 & 0.201 \\
LMC$\_$SC18& 0.182 & 0.178 & 0.173 & 0.170 \\
LMC$\_$SC19& 0.153 & 0.153 & 0.187 & 0.167 \\
LMC$\_$SC20& 0.132 & 0.137 & 0.142 & 0.163 \\
LMC$\_$SC21& 0.133 & 0.152 & 0.145 & 0.146 \\
\hline}

The zero points of our reddening map were derived based on previous
determinations in three lines-of-sight, two of them using OB-stars.
These determinations included determination of reddening around two LMC
star clusters: NGC1850 ($E(B-V)=0.15\pm0.05$~mag, based on {\it UBV}
photometry,  Lee 1995) and NGC1835 ($E(B-V)=0.13\pm0.03$~mag, based on
colors of RR Lyr stars, Walker 1993) and determination based on OB-stars
in the field of the eclipsing variable star HV2274 (Udalski \etal
1998c). All these zero points were consistent with our map to within a
few thousands of magnitude. 

We also checked the absolute calibration of our map comparing the
observed {\it I}-band magnitude of red clump stars with extinction free
magnitude determined from a few star clusters in the halo of the LMC
(Udalski 1998b). We additionally checked the value of extinction free
magnitude of red clump stars in the LMC by its new determination from
the field stars around the same clusters. Resulting value was consistent
to within 0.01~mag with star cluster red clump determination. The
calibration {\it via} extinction free magnitude of red clump stars gave
somewhat larger zero point of the $E(B-V)$ reddening -- by about
0.02~mag which we adopt as the error of our map.  The final $E(B-V)$
reddening in 84 lines-of-sight in the LMC is listed in Table~2.
Interstellar extinction in the {\it BVI} bands was calculated using the
standard extinction curve coefficients (\eg Schlegel \etal 1998):

$$ A_B=4.32\cdot E(B-V) $$
$$ A_V=3.24\cdot E(B-V) $$
$$ A_I=1.96\cdot E(B-V) $$

\Subsection{Astrometry} 

Equatorial coordinates of all candidates were calculated based on
transformation derived with the Digitized Sky Survey  images. Details of
procedure are described in Udalski \etal (1998b). About $2800-7400$
stars common in OGLE and DSS images (depending on stellar density of the
field) were used for transformation. Internal accuracy of the equatorial
coordinates is about 0.15 arcsec with possible systematic errors of the
DSS coordinate system up to 0.7 arcsec.

\Subsection{Fourier Parameters of Light Curve Decomposition}

For each object we derived Fourier parameters $R_{21}=A_2/A_1$ and
$\phi_{21}=\phi_2-2\phi_1$ of the Fourier series decomposition of
{\it I}-band light curve. $A_i$ and $\phi_i$ are the amplitudes and
phases of $(i-1)$  harmonic of the Fourier decomposition of light curve.
 Parameters $R_{21}$ and $\phi_{21}$ are often used for analyses of
pulsating variable stars and for discrimination between objects
pulsating in different modes.

We fitted the fifth order Fourier series to the magnitude scale {\it
I}-band light curve. In the case of objects with almost sinusoidal light
curve for which the first harmonic amplitude and phase were not
statistically significant, $R_{21}=0$ and $\phi_{21}$ is not defined. 

\begin{figure}[htb]
\hglue-0.5cm\psfig{figure=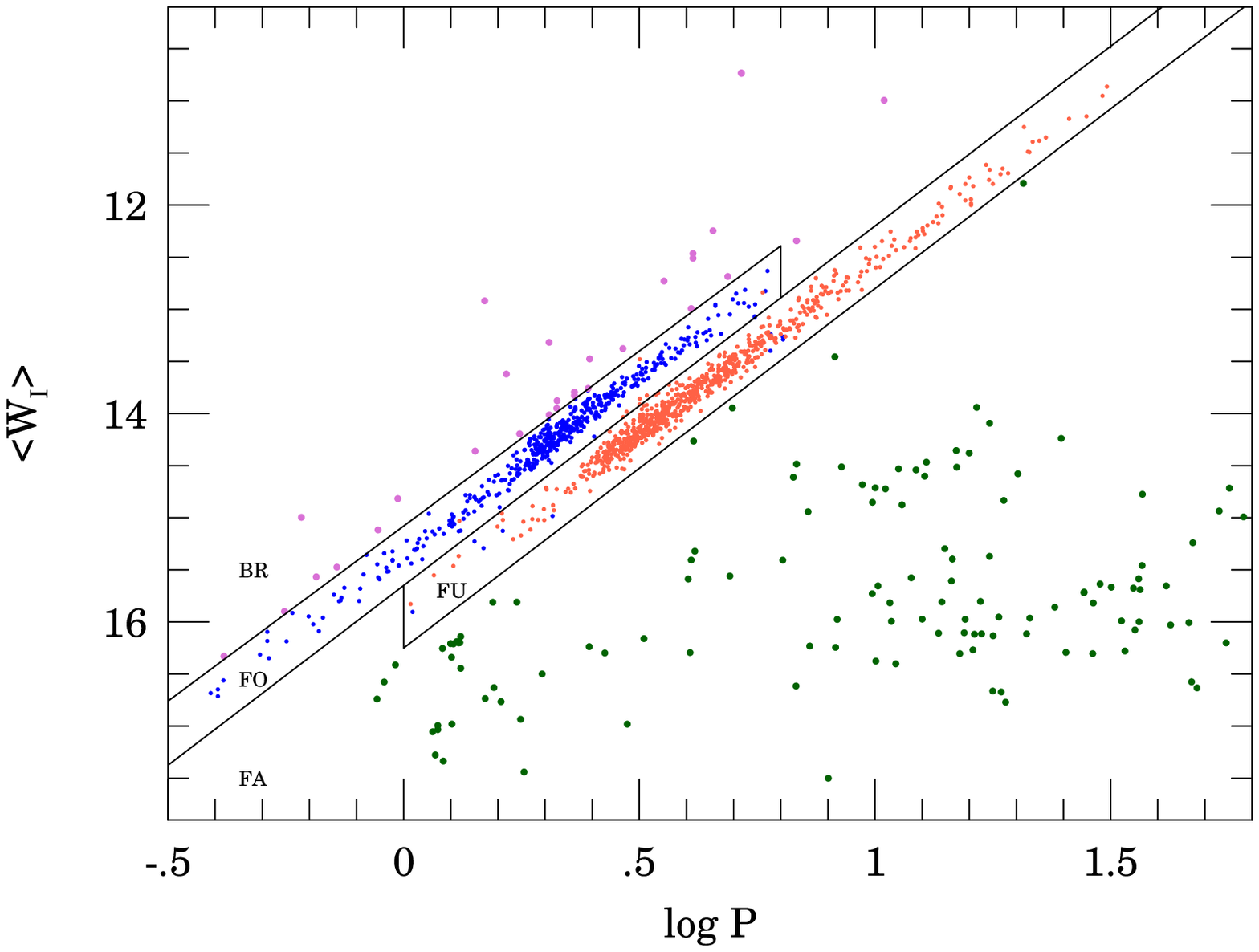,bbllx=30pt,bblly=50pt,bburx=510pt,bbury=410pt,width=12.5cm,clip=}
\vspace*{3pt}
\FigCap{Period-Luminosity relation for extinction insensitive index
$W_I$. Contours divide the diagram into sections where  fundamental
(FU) and first overtone mode (FO) classical Cepheids are found. Section
denoted by BR indicates region where objects were classified as brighter
than FO Cepheids and by FA -- as fainter than FU Cepheids. Small dots
mark positions of objects finally classified as FU and FO classical
Cepheids (light and dark dots, respectively). Larger dots -- BR (light
dots) and FA (dark dots) objects.}
\end{figure}

\Subsection{Classification}

Based on the Period-Luminosity ($P-L$) diagram constructed for the
extinction insensitive index $W_I$ we divided all objects into four
groups: classical Cepheids pulsating in the fundamental mode (FU),
classical Cepheids pulsating in the first overtone mode (FO), objects
brighter than FO mode Cepheids (BR) and objects fainter than FU mode
Cepheids (FA). Fig.~2 presents $P-L$ diagram for the $W_I$ index with
boundaries of these four regions.

Due to very good accuracy of photometry and features of the $W_I$ index,
which removes simultaneously effects of extinction and color dependence
of the Cepheid $P-L$ relation, the separation between the FU and FO
Cepheids is remarkable. Nevertheless, we also checked location of all
selected FU and FO Cepheids in the $R_{21}$ and $\phi_{21}$ \vs $\log P$
diagrams. It is well known that such diagrams allow to separate between
the FU and FO mode pulsators (\cf Alcock \etal 1999, Udalski \etal
1999a). Sequences for FU and FO Cepheids in both diagrams, in particular
$R_{21}$ \vs $\log P$, are well separated and in most cases
classification is straightforward. However, in a few period ranges  the
sequences almost overlap (for $0.6<\log P < 0.8$ in the $R_{21}$ \vs
$\log P$ diagram and $0.2<\log P<0.4$ and $\log P\approx0.75$ in the
$\phi_{21}$ \vs $\log P$ diagram). Therefore we checked light curves of
all objects located in these regions to confirm classification indicated
by position in the $P-L$ diagram. Also all objects located in opposite
mode sequences than indicated from $P-L$ position were inspected. In
about 20 cases the classification was changed. In eight cases those were
FU objects blended with other stars and therefore shifted to FO Cepheids
in the $P-L$ diagram. In eleven cases --  FO mode stars shifted to FU
objects in the {\it P--L} diagram of the $W_I$ index because of high
reddening, blends with blue stars etc. Fig.~3 presents the final
$R_{21}$ \vs $\log P$ and $\phi_{21}$ \vs $\log P$ diagrams for all
objects classified as FU and FO mode classical Cepheids. 

\begin{figure}[p]
\psfig{figure=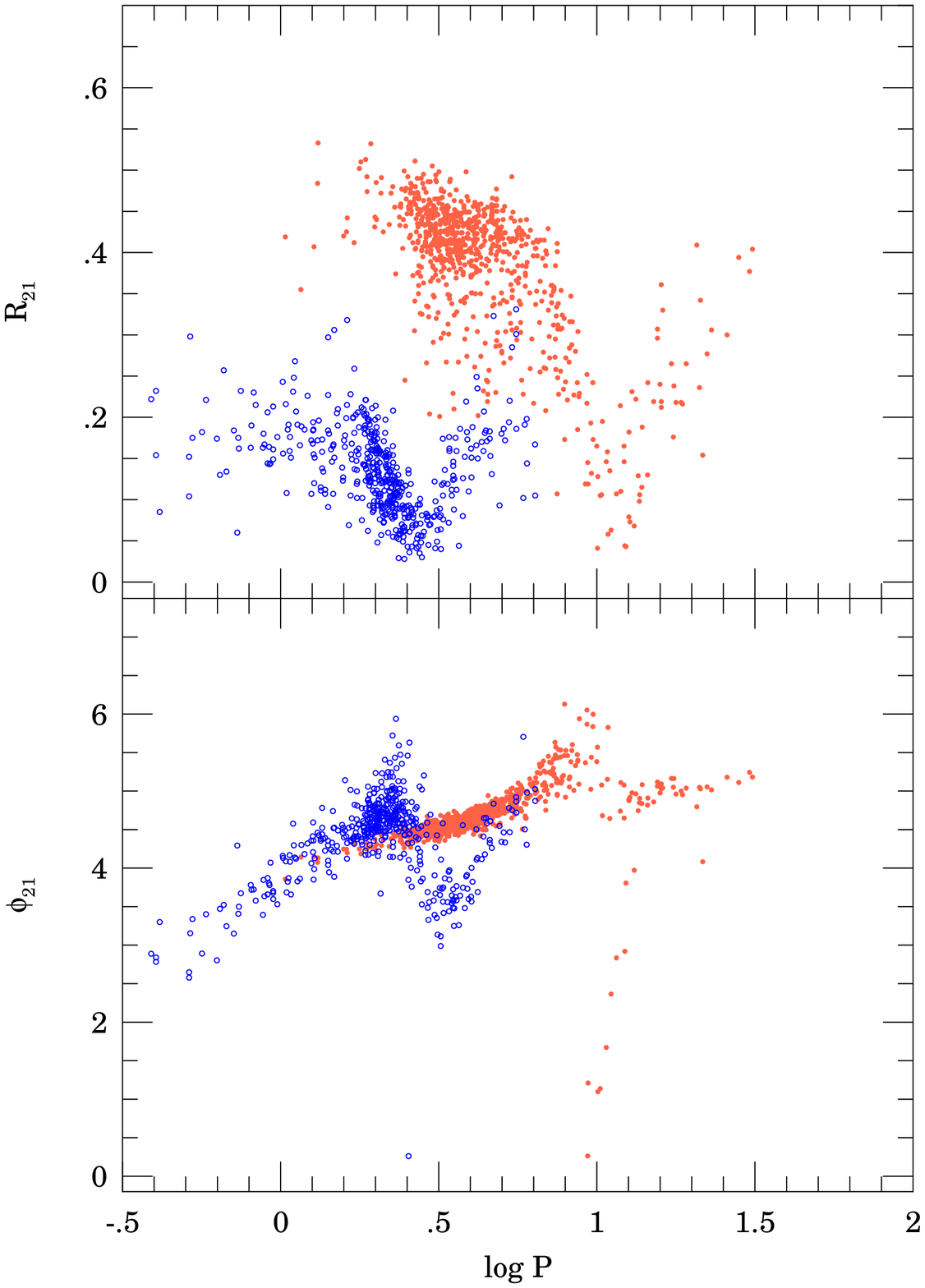,bbllx=30pt,bblly=40pt,bburx=505pt,bbury=705pt,width=12.5cm
,clip=}
\FigCap{${R_{21}}$ and ${\phi_{21}}$ \vs $\log P$ diagrams for
single-mode classical Cepheids from the LMC. Dark open circles indicate
positions of first overtone Cepheids while light dots positions of
fundamental mode pulsators.}
\end{figure}

\newpage

\Section{Catalog of Cepheids from the LMC}

1402 objects passed our selection criteria described in Section~3. Among
candidates for Cepheids in the LMC  a subsample of double-mode classical
Cepheids containing about 70 objects was extracted. These objects  will be
presented in a separate paper of this series. The remaining Cepheid
candidates are listed in Table~3.

The first column of Table~3 is the star identification: {\it
field$\_$name star$\_$num\-ber}. In the next columns the equatorial
coordinates, RA and DEC (J2000), period in days and moment of the zero
phase corresponding to maximum light are given. Then follow intensity
mean {\it IVB} photometry supplemented by extinction insensitive index
$W_I$ and the mean interstellar reddening in object direction. In the
next two columns Fourier parameters, $R_{21}$ and $\phi_{21}$, of the
light curve decomposition are listed. Finally, in the last column
classification of the object is provided.

Table~3 contains 1435 entries but only 1333 objects: 102 stars were
detected twice -- in overlapping parts of adjacent fields. Table~4
provides cross-identification of all such objects including for
completeness three double-mode Cepheids not presented in this paper and
not listed in Table~3.

The {\it I}-band light curves of all objects are presented in Appendices
A--U. The ordinate is phase with 0.0 value corresponding to maximum
light. Abscissa is the {\it I}-band  magnitude. The light curve is
repeated twice for clarity.

Finding charts ($60\arcs\times 60\arcs$ part of the {\it I}-band image)
are not presented in this paper but they are available in electronic
form from the OGLE Internet archive (see below).

We did not attempt to cross-identify our objects with the ones known
from literature. It would be a very difficult and time consuming task in
so dense stellar fields without precise coordinates and finding charts.
Because of very high completeness of the Catalog  (see Section~6)
-- practically all Cepheids in the observed region of the LMC have been
detected -- and precise {\it BVI} photometry the OGLE catalog is likely
to supersede much of the previous works. However, if necessary,
cross-identification with selected objects can be done with precise
coordinates and finding charts provided with the Catalog.

One should remember about the limit of the Catalog on the brighter, \ie
longer period object side because of saturation of the CCD detector
used. This limit corresponds to objects with period longer than $\log
P\approx1.5$, \ie longer than about 30~days.

\renewcommand{\TableFont}{\scriptsize}
%\vspace*{-25pt}
\setcounter{table}{3}
\MakeTableSep{
l@{\hspace{3pt}}
r@{\hspace{4pt}}
c@{\hspace{4pt}}
l@{\hspace{3pt}}
r@{\hspace{30pt}}
l@{\hspace{3pt}}
r@{\hspace{4pt}}
c@{\hspace{4pt}}
l@{\hspace{3pt}}
r}{12.5cm}{Cross-identification of stars detected in overlapping regions}
{
LMC$\_$SC1  & 275300  & $\leftrightarrow$ & LMC$\_$SC16  & 11573    & LMC$\_$SC12 & 210606  & $\leftrightarrow$ & LMC$\_$SC11  & 37982  \\
LMC$\_$SC1  & 290572  & $\leftrightarrow$ & LMC$\_$SC16  & 26258    & LMC$\_$SC13 & 12      & $\leftrightarrow$ & LMC$\_$SC12  & 51598  \\
LMC$\_$SC1  & 295698  & $\leftrightarrow$ & LMC$\_$SC16  & 31555    & LMC$\_$SC13 & 203493  & $\leftrightarrow$ & LMC$\_$SC12  & 210640 \\
LMC$\_$SC1  & 313151  & $\leftrightarrow$ & LMC$\_$SC16  & 48261    & LMC$\_$SC14 & 216182  & $\leftrightarrow$ & LMC$\_$SC12  & 47109  \\
LMC$\_$SC1  & 324986  & $\leftrightarrow$ & LMC$\_$SC16  & 57446    & LMC$\_$SC14 & 252799  & $\leftrightarrow$ & LMC$\_$SC13  & 35171  \\
LMC$\_$SC2  & 334104  & $\leftrightarrow$ & LMC$\_$SC1   & 31577    & LMC$\_$SC15 & 181509  & $\leftrightarrow$ & LMC$\_$SC14  & 11554  \\
LMC$\_$SC2  & 334077  & $\leftrightarrow$ & LMC$\_$SC1   & 31612    & LMC$\_$SC16 & 211276  & $\leftrightarrow$ & LMC$\_$SC17  & 11199  \\
LMC$\_$SC2  & 349919  & $\leftrightarrow$ & LMC$\_$SC1   & 44657    & LMC$\_$SC16 & 214736  & $\leftrightarrow$ & LMC$\_$SC17  & 15426  \\
LMC$\_$SC2  & 357856  & $\leftrightarrow$ & LMC$\_$SC1   & 51886    & LMC$\_$SC16 & 218499  & $\leftrightarrow$ & LMC$\_$SC17  & 19793  \\
LMC$\_$SC2  & 365487  & $\leftrightarrow$ & LMC$\_$SC1   & 59277    & LMC$\_$SC16 & 222451  & $\leftrightarrow$ & LMC$\_$SC17  & 24561  \\
LMC$\_$SC3  & 368185  & $\leftrightarrow$ & LMC$\_$SC2   & 39166    & LMC$\_$SC16 & 222497  & $\leftrightarrow$ & LMC$\_$SC17  & 24574  \\
LMC$\_$SC3  & 376576  & $\leftrightarrow$ & LMC$\_$SC2   & 47348    & LMC$\_$SC16 & 222476  & $\leftrightarrow$ & LMC$\_$SC17  & 24585  \\
LMC$\_$SC3  & 384972  & $\leftrightarrow$ & LMC$\_$SC2   & 55470    & LMC$\_$SC16 & 230207  & $\leftrightarrow$ & LMC$\_$SC17  & 33268  \\
LMC$\_$SC3  & 393065  & $\leftrightarrow$ & LMC$\_$SC2   & 63342    & LMC$\_$SC16 & 230281  & $\leftrightarrow$ & LMC$\_$SC17  & 33286  \\
LMC$\_$SC3  & 408692  & $\leftrightarrow$ & LMC$\_$SC2   & 78855    & LMC$\_$SC16 & 230222  & $\leftrightarrow$ & LMC$\_$SC17  & 33289  \\
LMC$\_$SC4  & 369748  & $\leftrightarrow$ & LMC$\_$SC3   & 5709     & LMC$\_$SC16 & 230285  & $\leftrightarrow$ & LMC$\_$SC17  & 33290  \\
LMC$\_$SC4  & 369698  & $\leftrightarrow$ & LMC$\_$SC3   & 5715     & LMC$\_$SC16 & 230224  & $\leftrightarrow$ & LMC$\_$SC17  & 33292  \\
LMC$\_$SC4  & 399359  & $\leftrightarrow$ & LMC$\_$SC3   & 35233    & LMC$\_$SC16 & 230228  & $\leftrightarrow$ & LMC$\_$SC17  & 33299  \\
LMC$\_$SC4  & 399429  & $\leftrightarrow$ & LMC$\_$SC3   & 35297    & LMC$\_$SC16 & 230273  & $\leftrightarrow$ & LMC$\_$SC17  & 33351  \\
LMC$\_$SC4  & 408738  & $\leftrightarrow$ & LMC$\_$SC3   & 44256    & LMC$\_$SC16 & 230290  & $\leftrightarrow$ & LMC$\_$SC17  & 33368  \\
LMC$\_$SC4  & 417847  & $\leftrightarrow$ & LMC$\_$SC3   & 53226    & LMC$\_$SC16 & 240480  & $\leftrightarrow$ & LMC$\_$SC17  & 45198  \\
LMC$\_$SC4  & 417864  & $\leftrightarrow$ & LMC$\_$SC3   & 53242    & LMC$\_$SC16 & 240497  & $\leftrightarrow$ & LMC$\_$SC17  & 45207  \\
LMC$\_$SC4  & 418294  & $\leftrightarrow$ & LMC$\_$SC3   & 53702    & LMC$\_$SC16 & 240460  & $\leftrightarrow$ & LMC$\_$SC17  & 45218  \\
LMC$\_$SC5  & 343036  & $\leftrightarrow$ & LMC$\_$SC4   & 13514    & LMC$\_$SC16 & 245458  & $\leftrightarrow$ & LMC$\_$SC17  & 50018  \\
LMC$\_$SC5  & 424993  & $\leftrightarrow$ & LMC$\_$SC4   & 98151    & LMC$\_$SC16 & 245478  & $\leftrightarrow$ & LMC$\_$SC17  & 50024  \\
LMC$\_$SC6  & 356421  & $\leftrightarrow$ & LMC$\_$SC21  & 193117   & LMC$\_$SC16 & 253780  & $\leftrightarrow$ & LMC$\_$SC17  & 59761  \\
LMC$\_$SC6  & 356421  & $\leftrightarrow$ & LMC$\_$SC5   & 6197     & LMC$\_$SC16 & 253794  & $\leftrightarrow$ & LMC$\_$SC17  & 59808  \\
LMC$\_$SC6  & 369970  & $\leftrightarrow$ & LMC$\_$SC5   & 19786    & LMC$\_$SC17 & 189196  & $\leftrightarrow$ & LMC$\_$SC18  & 6114   \\
LMC$\_$SC6  & 369993  & $\leftrightarrow$ & LMC$\_$SC5   & 19806    & LMC$\_$SC17 & 197292  & $\leftrightarrow$ & LMC$\_$SC18  & 17714  \\
LMC$\_$SC6  & 377026  & $\leftrightarrow$ & LMC$\_$SC5   & 26913    & LMC$\_$SC17 & 200401  & $\leftrightarrow$ & LMC$\_$SC18  & 17724  \\
LMC$\_$SC6  & 404601  & $\leftrightarrow$ & LMC$\_$SC5   & 58244    & LMC$\_$SC17 & 200430  & $\leftrightarrow$ & LMC$\_$SC18  & 20948  \\
LMC$\_$SC6  & 405017  & $\leftrightarrow$ & LMC$\_$SC5   & 67261    & LMC$\_$SC17 & 203767  & $\leftrightarrow$ & LMC$\_$SC18  & 25000  \\
LMC$\_$SC6  & 422324  & $\leftrightarrow$ & LMC$\_$SC5   & 75989    & LMC$\_$SC17 & 207480  & $\leftrightarrow$ & LMC$\_$SC18  & 25015  \\
LMC$\_$SC7  & 356873  & $\leftrightarrow$ & LMC$\_$SC6   & 27321    & LMC$\_$SC17 & 207506  & $\leftrightarrow$ & LMC$\_$SC18  & 25041  \\
LMC$\_$SC7  & 372083  & $\leftrightarrow$ & LMC$\_$SC6   & 40971    & LMC$\_$SC17 & 211310  & $\leftrightarrow$ & LMC$\_$SC18  & 29237  \\
LMC$\_$SC7  & 380269  & $\leftrightarrow$ & LMC$\_$SC6   & 49297    & LMC$\_$SC17 & 214843  & $\leftrightarrow$ & LMC$\_$SC18  & 33576  \\
LMC$\_$SC7  & 415723  & $\leftrightarrow$ & LMC$\_$SC6   & 85035    & LMC$\_$SC17 & 214859  & $\leftrightarrow$ & LMC$\_$SC18  & 33591  \\
LMC$\_$SC7  & 425296  & $\leftrightarrow$ & LMC$\_$SC6   & 86027    & LMC$\_$SC17 & 214860  & $\leftrightarrow$ & LMC$\_$SC18  & 33594  \\
LMC$\_$SC7  & 432869  & $\leftrightarrow$ & LMC$\_$SC6   & 102424   & LMC$\_$SC17 & 224169  & $\leftrightarrow$ & LMC$\_$SC18  & 45433  \\
LMC$\_$SC7  & 440072  & $\leftrightarrow$ & LMC$\_$SC6   & 102475   & LMC$\_$SC18 & 174802  & $\leftrightarrow$ & LMC$\_$SC19  & 18878  \\
LMC$\_$SC7  & 447509  & $\leftrightarrow$ & LMC$\_$SC6   & 118107   & LMC$\_$SC18 & 188926  & $\leftrightarrow$ & LMC$\_$SC19  & 38257  \\
LMC$\_$SC8  & 312191  & $\leftrightarrow$ & LMC$\_$SC7   & 55965    & LMC$\_$SC18 & 195674  & $\leftrightarrow$ & LMC$\_$SC19  & 44947  \\
LMC$\_$SC8  & 326025  & $\leftrightarrow$ & LMC$\_$SC7   & 79610    & LMC$\_$SC18 & 199069  & $\leftrightarrow$ & LMC$\_$SC19  & 48659  \\
LMC$\_$SC8  & 337497  & $\leftrightarrow$ & LMC$\_$SC7   & 86332    & LMC$\_$SC18 & 202349  & $\leftrightarrow$ & LMC$\_$SC19  & 48662  \\
LMC$\_$SC8  & 337546  & $\leftrightarrow$ & LMC$\_$SC7   & 93939    & LMC$\_$SC19 & 157749  & $\leftrightarrow$ & LMC$\_$SC20  & 21111  \\
LMC$\_$SC9  & 286128  & $\leftrightarrow$ & LMC$\_$SC8   & 10158    & LMC$\_$SC19 & 163667  & $\leftrightarrow$ & LMC$\_$SC20  & 28803  \\
LMC$\_$SC9  & 342082  & $\leftrightarrow$ & LMC$\_$SC8   & 52668    & LMC$\_$SC19 & 175567  & $\leftrightarrow$ & LMC$\_$SC20  & 43620  \\
LMC$\_$SC9  & 349881  & $\leftrightarrow$ & LMC$\_$SC8   & 64709    & LMC$\_$SC19 & 178247  & $\leftrightarrow$ & LMC$\_$SC20  & 47209  \\
LMC$\_$SC9  & 372259  & $\leftrightarrow$ & LMC$\_$SC8   & 76176    & LMC$\_$SC21 & 187856  & $\leftrightarrow$ & LMC$\_$SC5   & 16     \\
LMC$\_$SC9  & 372261  & $\leftrightarrow$ & LMC$\_$SC8   & 76179    & LMC$\_$SC21 & 187797  & $\leftrightarrow$ & LMC$\_$SC5   & 19     \\
LMC$\_$SC10 & 250322  & $\leftrightarrow$ & LMC$\_$SC9   & 45301    & LMC$\_$SC21 & 187853  & $\leftrightarrow$ & LMC$\_$SC5   & 63     \\
LMC$\_$SC10 & 256258  & $\leftrightarrow$ & LMC$\_$SC9   & 52800    & LMC$\_$SC21 & 193117  & $\leftrightarrow$ & LMC$\_$SC5   & 6197   \\
LMC$\_$SC11 & 306294  & $\leftrightarrow$ & LMC$\_$SC10  & 35605    & & & & & \\
}

\Section{Completeness of the Catalog}

Cepheids belong to brighter objects among stars in the OGLE photometric
databases. With large amplitude of light variations  they are relatively
easy to detect. Therefore, one can expect that completeness of our
Catalog is high.

Completeness of the Catalog can be estimated based on comparison of
number of objects detected in the overlapping regions between the
neighboring fields. 23 such regions exist between our fields  (Fig.~1)
allowing to perform 46 tests of pairing objects from a given and
adjacent fields. We analyzed full sample of our candidates including
detected double-mode  Cepheids to increase statistic. In total 217
objects from our full list (Table~3 plus double-mode Cepheid sample)
should be paired with counterparts in the overlapping field. We found
counterparts in 210 cases which yields the completeness of our sample
equal to 96.8\%. Thus, our tests indicate that the completeness of our
Catalog is indeed very high -- practically all Cepheids from the
observed fields have been detected.

The completeness is very likely to be even higher. The regions at the
field edge are biased in general by smaller number of observations due
to imperfections in telescope pointing. Because one of conditions which
a star had to fulfill to be searched for variability was 50
observations, this could lead to omission of some objects. Indeed,
counterparts of four from our seven unpaired objects had a number of
observations smaller than requested and they were not searched for
variability at all. We easily detected them as Cepheids when this
condition was  removed. Two of the  remaining objects were missed
because of severe blending with other bright stars leading to very noisy
light curves. The last unpaired object was a small amplitude, almost
sinusoidal shape variable and its counterpart was misclassified as an
ellipsoidal variable star.

Comparison of Cepheids from overlapping fields does not take into
account completeness of detection of stars by the OGLE data pipeline. It 
can be derived with artificial star tests. Although such tests have not
been performed yet for our LMC fields we can estimate it based on
results of tests for the SMC fields with similar  stellar density
(Udalski \etal 1998b). For objects as bright as Cepheids it was found to
be larger than 99\%. Thus, we may conclude that the total completeness
of our Catalog is $>96$\%.

\Section{Discussion}

The OGLE Catalog of Cepheids in the LMC provides an unique,
statistically complete  sample of these stars ideal for analyzing their
properties. The distribution of objects in the LMC is shown in Fig.~1.
Dots indicate positions of objects within observed fields. One can
easily notice that the distribution is not uniform within the galaxy.
Large fraction of objects is located in the south-eastern part of the
LMC: in the fields LMC$\_$SC16, LMC$\_$SC17 and LMC$\_$SC18. This region
must contain many younger stars and therefore Cepheids are more numerous
there than in other regions of the LMC. To make comparison  more
quantitative Table~5 lists for each field number of objects from the
Catalog, number of all stellar objects detected  in the field and number
of stars brighter than $I_0=17.5$~mag (approximately the limit of
brightness of classical Cepheids in the LMC).

\MakeTable{lrcc}{12.5cm}{Number of Cepheids and stars in the LMC fields.}
{\hline
\noalign{\vskip3pt}
\multicolumn{1}{c}{Field}&\multicolumn{1}{c}{$N_{Cep}$}& $N_{tot}$ & $N_{I_0<17.5}$   \\
\noalign{\vskip3pt}
\hline
\noalign{\vskip3pt}
LMC$\_$SC1 &  51~~ & 341528 & 27886 \\
LMC$\_$SC2 &  78~~ & 420043 & 31920 \\
LMC$\_$SC3 &  58~~ & 446233 & 34541 \\
LMC$\_$SC4 &  91~~ & 482991 & 37766 \\
LMC$\_$SC5 &  75~~ & 457988 & 37658 \\
LMC$\_$SC6 &  84~~ & 470171 & 38366 \\
LMC$\_$SC7 &  87~~ & 473469 & 39794 \\
LMC$\_$SC8 &  87~~ & 364573 & 35740 \\
LMC$\_$SC9 &  56~~ & 397307 & 28661 \\
LMC$\_$SC10&  39~~ & 292812 & 26460 \\
LMC$\_$SC11&  37~~ & 355750 & 24821 \\
LMC$\_$SC12&  19~~ & 215267 & 18562 \\
LMC$\_$SC13&  50~~ & 273347 & 20734 \\
LMC$\_$SC14&  50~~ & 264828 & 21195 \\
LMC$\_$SC15&  48~~ & 223749 & 16125 \\
LMC$\_$SC16& 145~~ & 269429 & 25061 \\
LMC$\_$SC17& 154~~ & 239856 & 19851 \\
LMC$\_$SC18&  91~~ & 212219 & 16783 \\
LMC$\_$SC19&  43~~ & 195590 & 15181 \\
LMC$\_$SC20&  49~~ & 209803 & 14116 \\
LMC$\_$SC21&  43~~ & 198314 & 14898 \\
\hline}

It should be also noted that positions of many Cepheids coincide with
areas of star clusters and it is very likely that many of them are star
cluster members. In the next paper of the series we will provide a full
list of Cepheids in the LMC clusters.

The Catalog provides ideal data for studying the Period--Luminosity and
Period--Luminosity--Color relations of classical Cepheids -- one of the
most important features of these stars. Detailed analysis of the $P-L$
and $P-L-C$ relations based on these data was presented in a separate
paper  (Udalski \etal 1999c).

\begin{figure}[p]
\hglue-0.5cm\psfig{figure=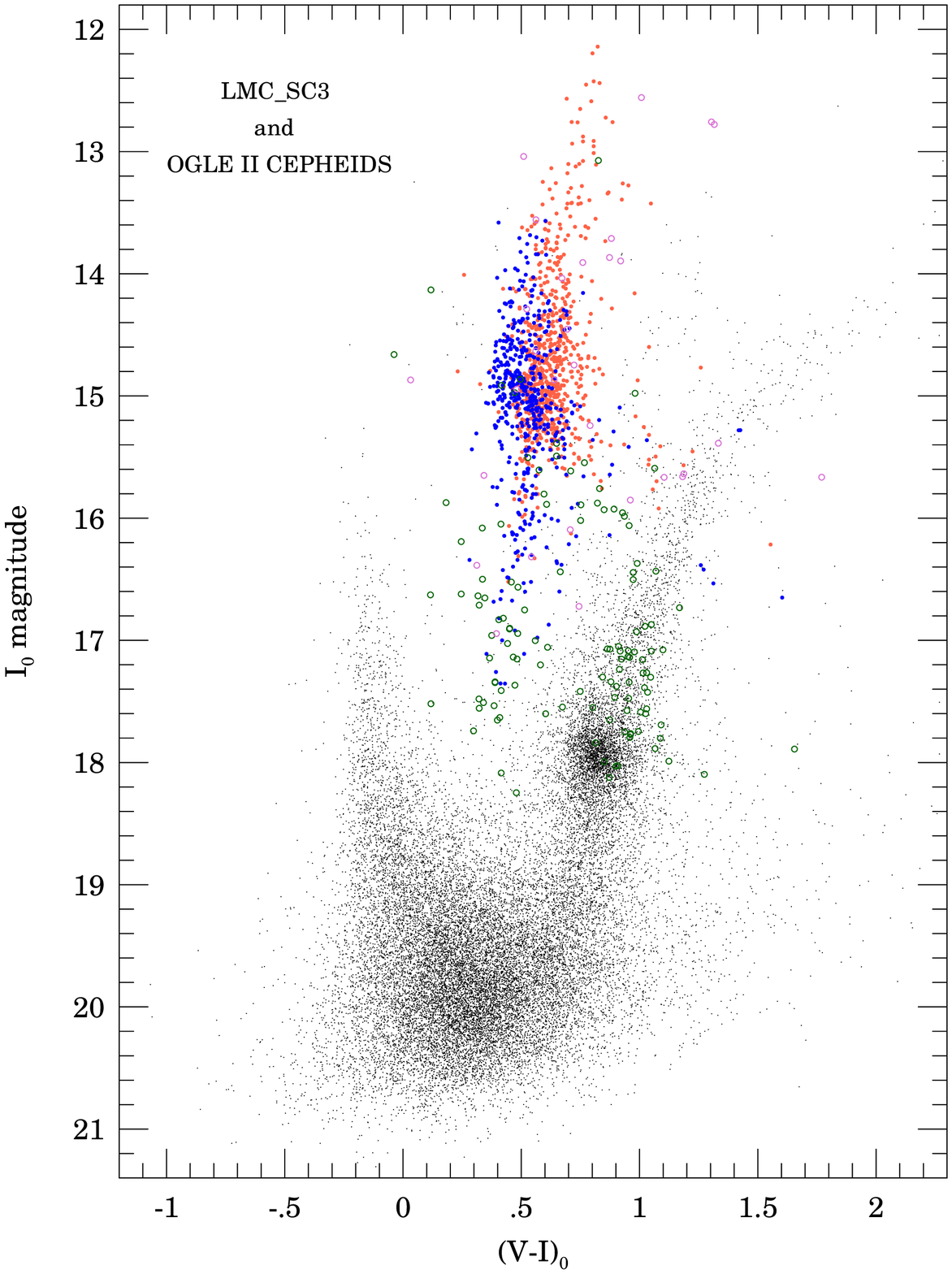,bbllx=35pt,bblly=50pt,bburx=550pt,bbury=730pt,width=13cm,clip=}
\FigCap{Color-magnitude diagram of subfield 2 of the LMC$\_$SC3
field. Only about 20\% of field  stars are plotted by tiny dots. Larger
dots show positions of FO and FU classical Cepheids (dark and light
dots, respectively). Dark and light open circles mark positions of
objects from the Catalog classified as FA and BR, respectively.}
\end{figure}

Fig.~4 shows the color-magnitude diagram (CMD) of subfield 2 area of the
LMC$\_$SC3 field corrected for the mean $E(B-V)=0.120$ reddening in this
direction (Table~2). Field stars from this field are plotted by tiny
dots. Larger dots indicate positions of classical Cepheids while open
circles positions of the remaining stars from our Catalog.

Based on location of stars in the CMD and $P-L$ $W_I$-index diagram
(Fig.~2) we may draw some conclusions on the objects classified as BR and FA
\ie brighter or fainter than FO and FU mode classical Cepheids. Brighter
objects (BR) are usually classical Cepheids, unresolved blends with
other star what shifts their magnitudes and colors and changes shape of
the light curve. We do not find among them any promising candidate for
second overtone mode classical Cepheid contrary to the SMC where large
sample of such stars was found (Udalski \etal 1999b).

\begin{figure}[htb]
\psfig{figure=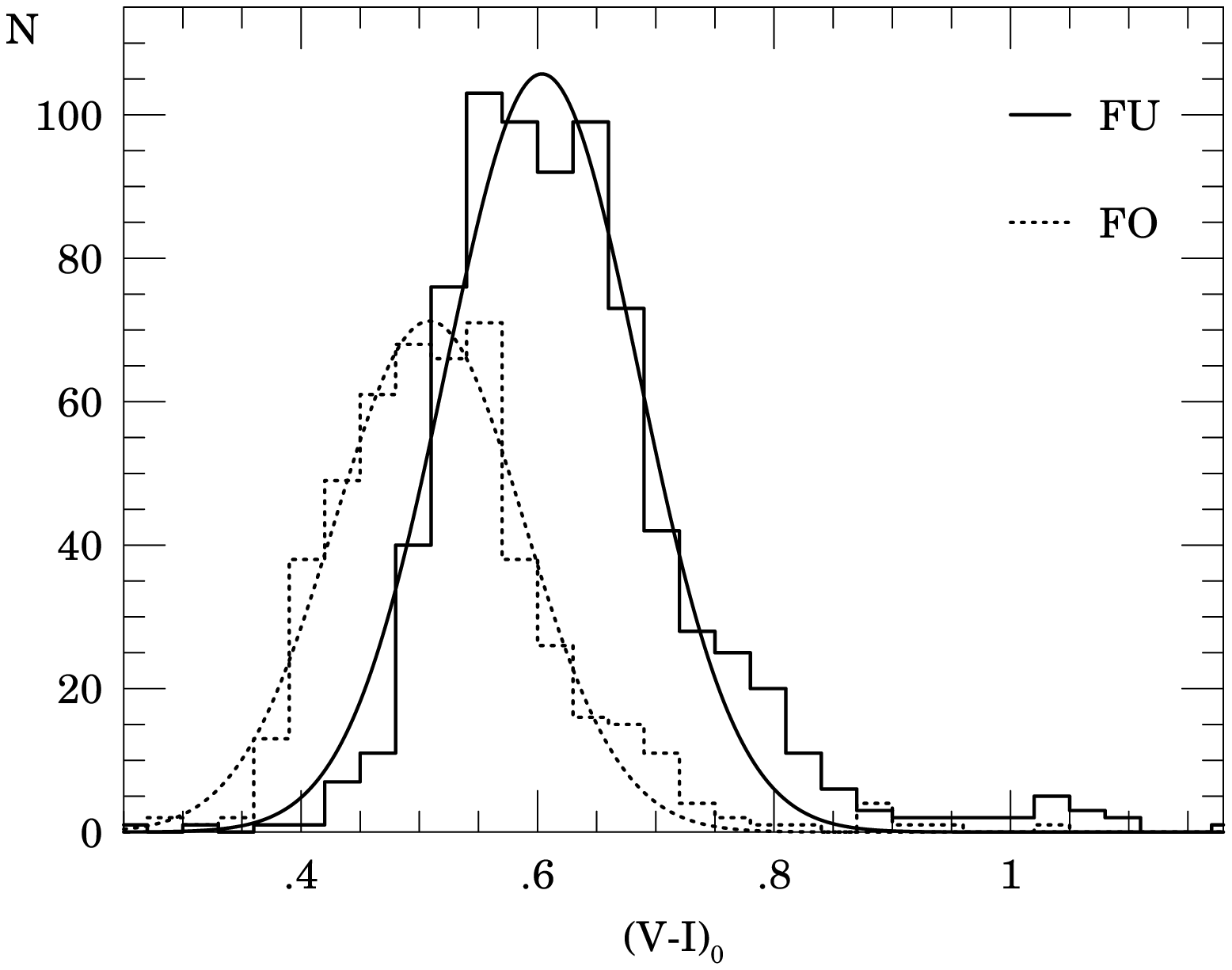,bbllx=50pt,bblly=50pt,bburx=505pt,bbury=410pt,width=12.5cm
,clip=}
\FigCap{Histograms of $(V-I)_0$ color distribution of single-mode 
Cepheids in the LMC. Solid line represents distribution of fundamental
mode pulsators, dotted line -- first overtone objects. The bins are
0.03~mag wide.}
\end{figure}

Among fainter objects two main classes can be distinguished. First one
consists of Population II Cepheids which are about 2~mag fainter than
classical Cepheids. They form a clear sequence below the $P-L$ relation
of classical  FU mode Cepheids (Fig.~2). The second group (about sixty
objects) contains stars with period in the range of $0.8 < \log P < 1.8$
and the mean $W_I\approx15.9$~mag. These objects are typically the red
giant branch stars and they do not form any noticeable $P-L$ relation.
Although their light curves resemble those of pulsating stars, it may
happen that their real variability is not related to pulsations. A few
shortest period stars from the FA group might be longer period RR~Lyr
stars located in front of the LMC (distance modulus about $1-1.5$~mag
smaller than that of the LMC). A few objects located in Fig.~2 close to
the boundary of FU mode Cepheids are highly reddened FU Cepheids or
Cepheids blended with blue stars.

Fig.~5 presents the distribution of color indices $V-I$ of classical FU
and FO mode Cepheids. The mean $(V-I)_0$ color and its dispersion are
equal to $(0.604, 0.08)$ and $(0.509, 0.08)$ for  the FU and FO mode
Cepheids in the LMC, respectively. The distribution can be well
approximated with a Gaussian but an excess of red objects is clearly
seen for both types of Cepheids. It is partially caused  by Cepheids
reddened more than the mean correction applied. This effect is also seen
in Fig.~4 for field stars -- the red clump is  slightly elongated in the
direction of reddening.  Another reason of excess of red objects is
bending of  the instability strip in the CMD diagram for brighter
(redder) objects.

\begin{figure}[htb]
\psfig{figure=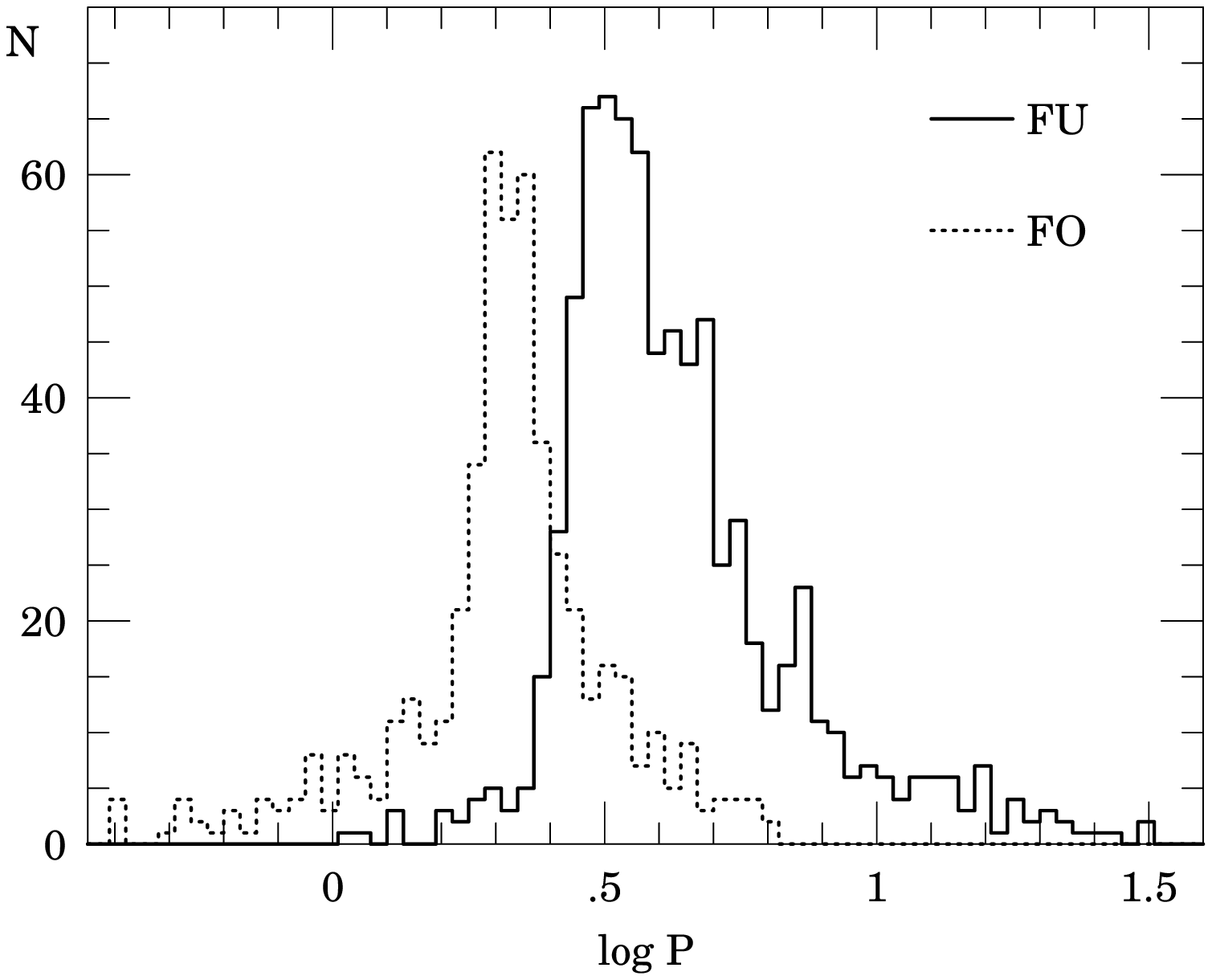,bbllx=50pt,bblly=50pt,bburx=505pt,bbury=410pt,width=12.5cm
,clip=}
\FigCap{Histograms of $\log P$ distribution of single-mode  Cepheids in
the LMC. Solid line represents distribution of fundamental mode
pulsators, dotted line -- first overtone objects. The bins are 0.03
wide in $\log P$.}
\end{figure}

Fig.~6 shows distribution of periods of FU and FO mode classical
Cepheids in the LMC. Typical period of the FU mode Cepheid in the LMC is
about 3.2~days while for the FO mode objects 2.1~days. The fundamental
mode Cepheid period distribution has a long tail toward long period
objects and the number of Cepheids with period shorter than 2.3~days
falls rapidly to zero. The longest period of the FO mode Cepheids is
about 6.4~days while the shortest periods are of about 0.4~day.

Different distribution of periods and different metallicity of Cepheids
in the LMC and SMC (${\rm [Fe/H]}=-0.3$~dex, and $-0.7$~dex for the LMC and SMC,
respectively, Luck \etal 1998) result in somewhat different diagrams of
Fourier parameters of light curve decomposition (\cf Fig.~3 and Fig.~2
of Udalski \etal 1999a). Detailed analysis of these diagrams may provide
information on dependence of the shape of Cepheid light curve on
metallicity.

\begin{figure}[htb]
\hglue1.5cm\psfig{figure=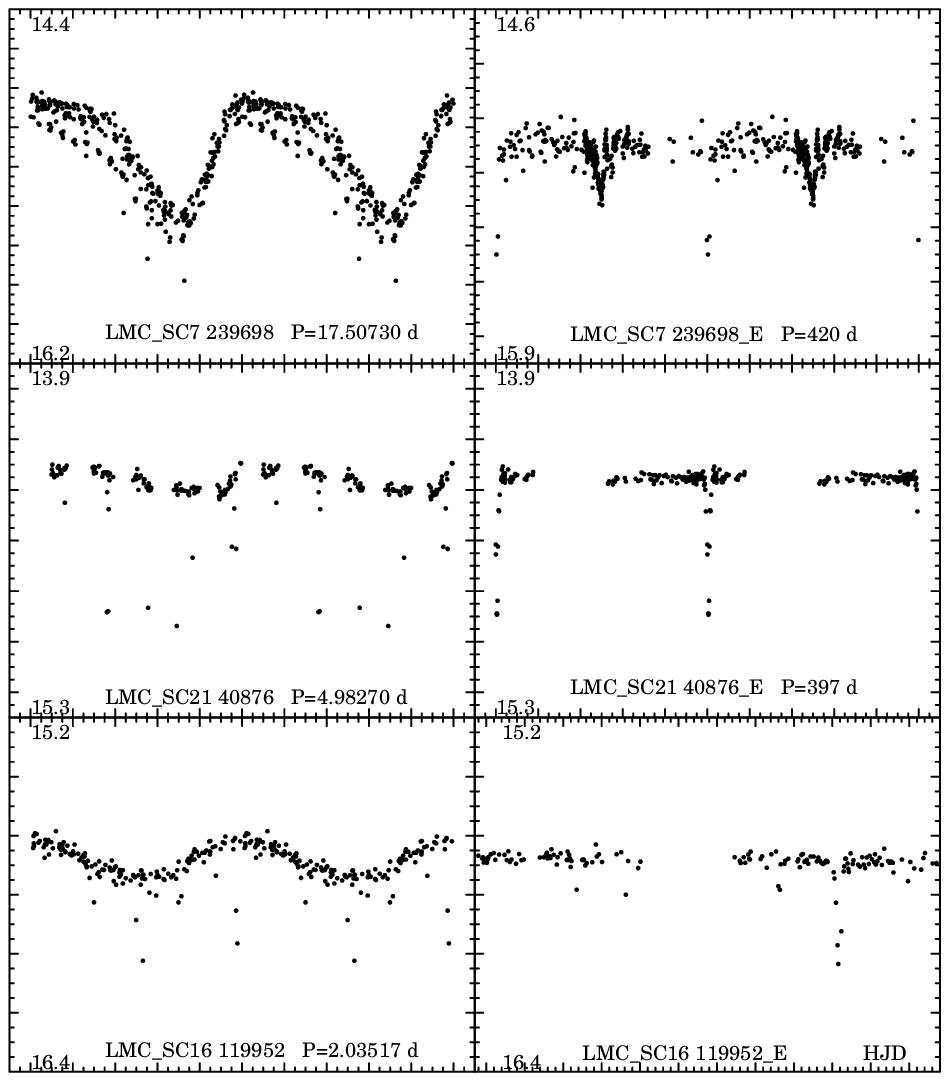,bbllx=125pt,bblly=275pt,bburx=400pt,bbury=590pt,width=9cm
,clip=}
\vspace*{3pt}
\FigCap{Light curves of three eclipsing  binary systems with Cepheid
component. In the left panel observations are folded with the Cepheid
period. In the right panel eclipsing light curve is shown after
subtracting the brightness of Cepheid component. Abscissa is the {\it
I}-band magnitude and the ordinate -- phase with 0.0 value corresponding
to maximum brightness in the left panels and middle of deeper eclipse in
two upper right panels. For LMC$\_$SC16 119952 the ordinate of the right
panel is in HJD with larger ticks separated by 50~days and the most
right large tick equal to ${\rm HJD}=2451300$.}
\end{figure}

Among many individual objects which show full variety of Cepheid
behaviors (\eg many "bump" Cepheids) three objects require a special
attention. They are the binary, eclipsing systems including Cepheid as a
component: LMC$\_$SC7 239698, LMC$\_$SC21 40876 and LMC$\_$SC16 119952.
The first two of these objects were already found by the MACHO group
(Welch \etal 1999). Such systems are very important because their
precise photometry and spectroscopy can provide accurate information on
sizes and masses of the components. LMC$\_$SC7 239698 is a Population II
Cepheid while the remaining stars are classical Cepheids. LMC$\_$SC21
40876 is a FU mode pulsator. LMC$\_$SC16 119952 pulsates in the FO mode.

Fig.~7 shows the light curves of these three Cepheids. In the left panel
the light curve phased with the Cepheid period is presented. In the
right panel the light curve with subtracted Cepheid variability (by
approximation of its light curve with Fourier series) is displayed.  For
LMC$\_$SC7 239698 and LMC$\_$SC21 40876 observations are  folded with
the eclipsing period. It should be noted that the eclipsing periods are
very preliminary because the binary systems containing Cepheids are very
wide and only 2--3 eclipses were observed during the entire period of
the OGLE observations. They will be refined after the next observing
seasons. For LMC$\_$SC16 119952 only one clear eclipse was observed,
thus we present its light curve in day units along the ordinate.

We also draw attention to another interesting object -- LMC$\_$SC6
330185. Its variations are consistent with a sum of light of two 
Cepheids either being optical blend or physically related.  Both
components are the FO mode pulsators with periods 2.48092 and
1.96376~days. This object was already presented by the MACHO group
(Alcock \etal 1995). 

The Catalog of Cepheids from the LMC is available now to the
astronomical community from the OGLE Internet archive:
\begin{center}
{\it http://www.astrouw.edu.pl/\~{}ftp/ogle} \\
{\it ftp://sirius.astrouw.edu.pl/ogle/ogle2/var$\_$stars/lmc/cep/catalog/}\\
\end{center}
or its US mirror
\begin{center}
{\it http://www.astro.princeton.edu/\~{}ogle}\\
{\it ftp://astro.princeton.edu/ogle/ogle2/var$\_$stars/lmc/cep/catalog/}\\
\end{center}

The data include the mean photometry,  individual {\it BVI} observations
of all objects and finding charts. We plan to update the Catalog in the
future when more observations are collected. We would also appreciate
information on any errors in the Catalog which are unavoidable in so
large data set.

\Acknow{We would like to thank Prof.\ Bohdan Paczy\'nski for many
discussions and important suggestions. We thank Dr.\ K.~Z.\ Stanek for
valuable comments and remarks on the paper. We are also grateful to Dr.\
D.\ Welch for pointing out wrong period of LMC$\_$SC16 119952 eclipsing
Cepheid in the original version of the paper. The paper was partly
supported by  the Polish KBN grants 2P03D00814 to  A.~Udalski and
2P03D00916 to M.~Szyma{\'n}ski. Partial support for the OGLE  project
was provided with the NSF  grant AST-9530478 and AST-9820314 to
B.~Paczy\'nski. We  acknowledge usage of The Digitized Sky Survey which
was produced at the Space  Telescope Science Institute based on
photographic data obtained using The UK  Schmidt Telescope, operated by
the Royal Observatory Edinburgh.}

\end{document}